\documentclass{article}
\usepackage{spconf,amsmath,graphicx, booktabs, xcolor, hyperref}

\title{FlowGrad: Using Motion for Visual Sound Source Localization}
\name{$\begin{array}{cc} 
\mbox{Rajsuryan Singh $^{1}$,
Pablo Zinemanas$^{1}$, 
Xavier Serra$^{1}$, 
Juan Pablo Bello$^{2}$, 
Magdalena Fuentes$^{2, 3}$}\end{array}$ }
\address{$^1$ MTG, Universitat Pompeu Fabra, Barcelona, Spain\\\
$^2$ MARL, New York University, New York, USA\\
$^3$ IDM, New York University, New York, USA}

\begin{document}
%
\maketitle
\begin{abstract}
Most recent work in visual sound source localization relies on semantic audio-visual representations learned in a self-supervised manner and, by design, excludes temporal information present in videos. While it proves to be effective for widely used benchmark datasets, the method falls short for challenging scenarios like urban traffic. This work introduces temporal context into the state-of-the-art methods for sound source localization in urban scenes using optical flow to encode motion information. An analysis of the strengths and weaknesses of our methods helps us better understand the problem of visual sound source localization and sheds light on open challenges for audio-visual scene understanding. The code and pretrained models are publicly available at \url{https://github.com/rrrajjjj/flowgrad}

\end{abstract}
\section{Introduction}
\label{sec:intro}

Vision and audition are complementary sources of information and their effective integration, i.e. the ability to localize sounds and connect them to visual objects, enables a rich understanding of a dynamic environment. Early attempts at modeling audio-visual perception exploited the synchrony between audio and visual events, e.g. lip movements aligned to speech, with probabilistic models \cite{hershey1999audio, fisher2000learning}, and canonical correlation analysis \cite{kidron2005pixels}. With recent advances in deep learning, especially in computer vision, the field has pivoted to deep-neural-network-based methods. A notable difference between the two approaches is the shift from using the temporal correlation between audio and video to the semantic similarity between them as the primary source of information for localization. This has happened to the extent that most state-of-the-art methods, except for a very few examples \cite{afouras2020self, zhao2019sound}, completely disregard the temporal context available in videos \cite{arandjelovic2017look, arandjelovic2018objects, senocak2018learning, oya2020we, chen2021localizing}. These methods focus on learning semantic auditory and visual representations in a self-supervised manner that enables sound source localization (SSL) via the similarity between audio and visual embeddings. This approach has been effective for the widely used benchmark datasets\cite{arandjelovic2017look, arandjelovic2018objects, senocak2018learning, oya2020we, chen2021localizing}, however recent work by Wu et al. has raised questions about the generalizability of these methods beyond these datasets \cite{wu2022listen}. They further point out the strong biases present in these benchmarks and demonstrate that the methods developed on these datasets fail to generalize to urban scenes.

Urban scene understanding has many potential applications in various sectors, including assistive devices for the hard-of-hearing, traffic monitoring, and autonomous driving. However, visual sound source localization (VSSL) in urban scenes is a challenging task, and state-of-the-art methods are not sufficient \cite{wu2022listen}. Benchmark datasets for VSSL, such as VGG-SS and Flickr, typically have only one sound source per image, whereas urban scenes often have multiple agents that may or may not be producing sounds. 
To address this issue, we investigate the use of temporal context in our approach.

We test our methods on Urbansas dataset \cite{fuentes2022urbansas}, which is an audio-visual dataset for detecting sound events in urban environments. We only use Urbansas for evaluation because other VSSL benchmarks have a bias towards static sound sources in the center of the image, making the inclusion of motion information unnecessary, and RCGrad has already been evaluated on other VSSL benchmarks in \cite{wu2022listen}. Our baseline model for Urbansas is RCGrad \cite{wu2022listen}, which is  the state-of-the-art.

We propose the use of optical flow as a means to incorporate temporal information and we explore hard-coded as well as learning-based algorithms to combine it with RCGrad. First, we use optical flow as a heuristic to filter stationary objects from the predictions of RCGrad and observe a significant improvement in localization performance, especially in curbing false
positives. Further, we add optical flow as a feature to the neural network in two ways: i) we add optical flow as an additional channel into the vision encoder, and ii) we train a separate optical flow encoder within the RCGrad framework. 

\begin{figure*}[ht!]
   \centering
   \includegraphics[width=0.9\linewidth]{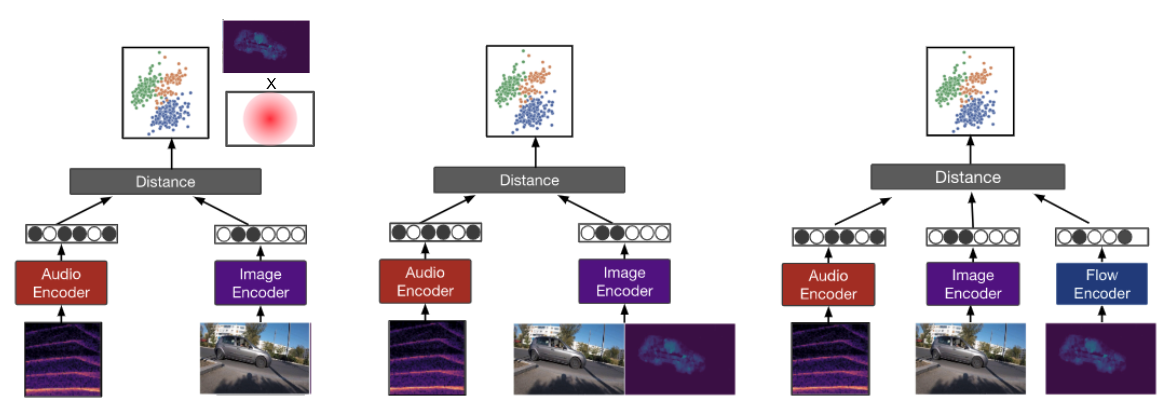}
   \caption{Left: FlowGrad-H, element-wise multiplication of the RCGrad predictions with optical flow. Center: FlowGrad-IC, optical flow as an extra channel in the image encoder. Right: FlowGrad-EN, optical flow added through a third flow encoder.  }
   \label{fig:models}
\end{figure*}

\section{Method}
\label{sec:method}

\subsection{RCGrad}
\label{ssec:rcgrad}

RCGrad \cite{wu2022listen} uses resnet-18 as the audio as well as the vision encoder. The vision encoder is pretrained on Imagenet while the audio encoder is randomly initialized. The model is then trained with a contrastive loss on VGG-Sound \cite{chen2020vggsound}. Each training example is a randomly selected image from a 5-second video along with the corresponding audio. As is standard in the literature, the model uses separate audio and vision encoders optimized with audio-visual correspondence as the training objective. Localization is done using a modified version of Grad-CAM \cite{selvaraju2017grad} wherein instead of back-propagating class labels, the audio embedding is back-propagated through the vision subnetwork to generate localization maps.

\subsection{FlowGrad: Incorporating temporal context}
\label{ssec:temporal_context}

\noindent \textbf{Optical Flow as a heuristic.} In the context of urban scene understanding, one major limitation of RCGrad is the attribution of sounds to parked vehicles. Since the representations are purely semantic and there is no temporal context, the model cannot distinguish between stationary and moving vehicles. As a result, parked vehicles often end up as false positives diminishing the performance. Optical flow, on the other hand, only has motion information. Anything that moves, be it vehicles, pedestrians, or tree leaves, have high values of activation. Hence, optical flow and RCGrad have complementary strengths that can be leveraged by taking an intersection of objects that have high activations for both. We execute this idea by simply doing an element-wise multiplication of the RCGrad predictions with the optical flow. This suppresses objects that are either not moving or that are moving but are not sounding, leaving us with sounding vehicles. We call this model \textit{FlowGrad-H}, depicted on the left of Figure \ref{fig:models}.

\noindent \textbf{Optical Flow as an image channel.} As effective as heuristics can prove to be, they are often rigid, brittle, and prone to a lack of generalizability. In an attempt to move away from the naive use of optical flow as a filter and towards using it to imbue the representations with temporality, we include it as an image channel. Here, the model can, at least in principle, take the motion information into account while making predictions, instead of motion being used as a filter post-hoc. The relationship between motion and sounds can hence be learned. To do so, we extended RCGrad to take in 4 channels (RGB and optical flow) as the input to the image encoder (see center of Figure \ref{fig:models}). We initialize the model with the pre-trained version of RC-Grad, and for the weights of the optical flow channel we used the average of the weights of the RGB channels. We train the model on the unlabeled portion of Urbansas using contrastive loss. Following \cite{wu2022listen}, during training, we fit the model with a frame along with a 5-second audio clip around it, plus the corresponding optical flow calculated between consecutive frames at 8fps as an additional channel. This model is \textit{FlowGrad-IC}.

\noindent \textbf{Optical Flow encoder.} In the above-mentioned method, the 4 channels of the image encoder are pooled in very early layers of the network. This may result in shallow integration of motion information. Moreover, since the model was initialized with weights pre-trained on audio and images, simply discarding the additional optical flow channel provides a trivial solution for minimizing the loss. To avoid this, we added a separate flow encoder with the same Resnet-18 architecture as the image and the audio encoders to RCGrad (see right of Figure \ref{fig:models}). We initialized the weights as the average of RGB channels of the vision encoder, and modified the training loss to be the sum of all pairwise losses (audio-image, image-flow, and audio-flow). The localization is then done by backpropagating the audio embeddings through the image as well as the flow encoder to generate two localization maps. These maps are then multiplied element-wise to give the final localization map. We call this model \textit{FlowGrad-EN}.

\section{Experimental design}
\label{sec:exp_design}


\noindent \textbf{The Urbansas dataset} is an audio-visual dataset developed for studying the detection and localization of sounding vehicles in the wild \cite{fuentes2022urbansas}. The dataset consists of labeled and unlabeled videos of urban traffic with stereo audio, to a total of 15 hours of video out of which 3 hours have been manually annotated, with both audio events and video bounding-box annotations, for sound event detection and source localization. We train on the unlabelled videos following the training protocol from \cite{wu2022listen} and we evaluated our models on the annotated portion of the dataset. For evaluation, we only consider frames that have both audio and video annotations and where the sounding vehicle is visible and identifiable giving us 5704 annotated image-audio pairs.  


\noindent \textbf{Baselines.} We employ three baselines: i) \textit{RCGrad} \cite{wu2022listen}, which is the current state-of-the-art localization method in Urbansas; and ii) a vision-only object recognition topline method with temporal and class filtering (\textit{vision-only+CF+TF}), which is a strong reference with temporal integration and information about the classes present but not sound
; and iii) an \textit{optical flow baseline}, which helps us understand how much of the data can be explained by motion only. 
We have replicated the results of RCGrad \cite{wu2022listen} using the pre-trained models from the official repository. 

For the vision-only+CF+TF baseline we use a pre-trained YOLOR object detection model \cite{wang2021you}. This model has been trained to predict bounding boxes around objects on the MS-COCO dataset \cite{lin2014microsoft}, which is a large-scale dataset with just under a million annotated objects, nearly 10\% of which correspond to vehicles. We use the pretrained \textit{yolor\_p6} model weights for inference, and we filter the results to the four vehicle classes present in Urbansas - car, motorcycle, bus, and truck. Further, we apply motion-based filtering. For each pair of consecutive frames (f and f+1), if a bounding box in f has an IoU greater than 0.95 with one in f+1, both the bounding boxes are discarded. This ensures that stationary objects are filtered out in the final predictions. For the optical flow baseline, we use the normalized optical flow directly as predictions, without any semantic filtering. This means that we consider that anything that is moving is producing sound. This method serves to demonstrate the correspondence, or a lack thereof, between moving and sounding objects. 

\noindent \textbf{Optical Flow.} The optical flow is calculated using the Gunnar Farneback algorithm \cite{farneback2003two}. Images are sampled at 8 frames-per-second, converted to grayscale, and dense optical flow is estimated between the current and the next frame using the OpenCV implementation of the algorithm. 

\noindent \textbf{Metrics}. Following \cite{wu2022listen}, the localization maps are min-max normalized and we use consensus intersection over union (cIoU) and area under the curve (AUC) as performance metrics as in the literature \cite{arandjelovic2018objects,senocak2018learning, chen2021localizing,  wu2022listen}. We binarize  localization maps with a threshold of 0.5 to calculate the cIoU. The AUC is calculated for cIoUs at different thersholds.

\section{Results and discussion}
\label{sec:discussion}

\begin{figure*}[ht!]
   \centering
   \includegraphics[width=\linewidth]{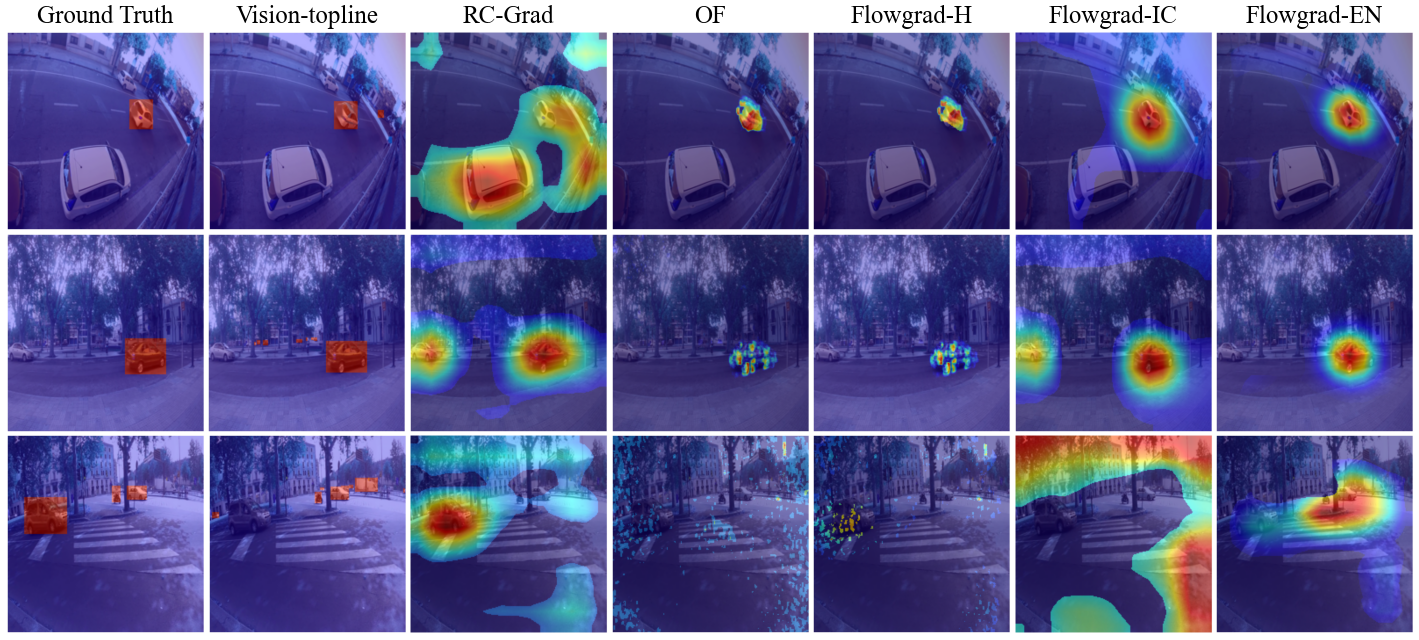}
   \caption{{Predictions of the baselines and the proposed models on selected examples. Optical flow proves to be effective in the first two examples but sounding vehicles parked at traffic signals are a limitation of the method as shown in the bottom row.}}
   \label{fig:main_result}
\end{figure*}

Results are presented in Table \ref{tab:results}. The first observation is that the vision topline performs  considerably better than the other methods.
 This is because YOLOR is a supervised object detection model that has information about classes and predicts precise bounding boxes around vehicles whereas the other models produce coarser and less precise heatmaps (see Figure \ref{fig:main_result}), scoring lower in the IoU. The ground truth annotations are also bounding boxes generated using an object detection model \cite{fuentes2022urbansas} and this congruence between the ground truth and the predictions further inflates the IoU. This combined with motion-based filtering gives us a very strong supervised reference to pit our self-supervised models against. 

\begin{table}[h]
    \centering
    \begin{tabular}{l|c|c} \toprule
    \textit{model} & \textit{IoU ($\tau=0.5$)} & \textit{AUC} \\
     \midrule

    \textbf{Vision-only+CF+TF (topline)} & \textbf{0.68} & \textbf{0.51} \\
    Optical flow only (baseline) & 0.33 & 0.23\\
    \midrule
    RCGrad \cite{wu2022listen} (sota)& 0.16 & 0.13\\
    \midrule
    \textbf{FlowGrad-H} & \textbf{0.50} & \textbf{0.30} \\ 
    FlowGrad-IC & 0.26 & 0.18\\ 
    FlowGrad-EN & 0.37 & 0.23\\ 
     \bottomrule
    \end{tabular}
    \caption{IoU and AUC results for the different models.}
    \label{tab:results}
\end{table}

All models that use motion information outperform RCGrad, 
since predictions of stationary vehicles are eliminated 
overcoming RCGrad's major limitation. Using thresholded optical flow directly as localization maps  outperforms vanilla RCGrad,
which suggests that there is a high correlation between motion and sound in Urbansas. As can be seen in the first two rows of Figure \ref{fig:main_result}, the optical flow baseline produces more precise localization heatmaps around moving vehicles, ignoring those that are parked, while the predictions of RCGrad focus on any visible car, including parked vehicles which are silent and hence are false positives. 

Looking at the results in Table \ref{tab:results}, we conclude that motion alone is not enough to explain sounding objects in urban settings, as the integration of motion; sound and semantics leads to the best performing unsupervised systems (FlowGrad-H and FlowGrad-EN). The best way of combining optical flow with the deep learning model seems to be as a post-processing heuristic (FlowGrad-H), followed by adding a flow encoder (FlowGrad-EN), and lastly, adding optical flow as an extra channel to the image encoder (FlowGrad-IC). A heuristic performing better than learning based methods is counterintuitive but it's been shown that if the dataset has strong biases, even trivial heuristics like a big-enough bounding box located in the middle of the image perform similar (and sometimes outperform) state-of-the-art methods \cite{wu2022listen} suggesting a strong sound-motion correspondence bias in Urbansas. With the integration of flow, the model is able to distinguish the parked vehicle (see Figure \ref{fig:main_result}), and the localization maps are for the most part less diffused and this stringency is likely to contribute to the increased IoU numbers due to a decrease in the overall area of union. By the same token, the size of the predicted masks may also, at least in part, explain why FlowGrad-EN does not perform as well as the naive use of optical flow as a heuristic (FlowGrad-H). Optical flow generates very precise masks around objects minimizing the area of union and hence increasing the IoU while this method still produces diffused localization maps. This opens up the question once more, as discussed in \cite{wu2022listen}, of whether bounding boxes combined with IoU are in fact a good way to evaluate localization models.

After examining the performance of our models on different scenes in Urbansas, we found that incorporating optical flow has its limitations and may even decrease performance in certain scenarios, such as those with shaky cameras. In cases where sound and motion do not correspond well, such as when parked vehicles produce sound, but are ignored by motion models, it is difficult for semantics or motion alone to describe the scene accurately. Therefore, we may need to use complementary information such as spatial sound or reasoning. Another approach we could explore to improve performance in these scenarios is to extend the temporal context window used in optical flow calculations. Currently, we use a short context window of 0.125 seconds, but increasing it to 5 seconds may provide the necessary information to attribute sounds to temporarily stationary vehicles. One way to do this is to aggregate optical flow over a 5-second window, similar to action recognition strategies, and use the resulting stack of optical flow as a feature. Alternatively, we could average the optical flow across the time window, as done in \cite{afouras2020self}.

Trees, pedestrians, and other moving objects are also exceptions to the assumption.  Moving tree leaves can often have high optical flow, but they have no contribution to the sounds whatsoever. However, in contrast to the previous example, using optical flow along with semantics and sound (as in FlowGrad) is a simple fix to this issue as the RCGrad predictions generally have very low activations for trees if the sounding object is an engine. The case with pedestrians is not as straightforward as it is for trees. They have characteristic sounds associated with them that are clearly audible, especially if they are close to the microphone. Most models we use for sound source localization (and certainly the ones investigated in this work) are class-agnostic and are trained in a self-supervised manner without any class labels. So RCGrad localizes pedestrians as sound sources as we have observed in some cases. Pedestrians also have high optical flow and hence cannot be filtered out by either method or a combination thereof. Since pedestrians are not labeled in the Urbansas dataset, they are evaluated as false positives. However, we think this is a limitation of the dataset rather than the method, and we will extend Urbansas' annotations in future work. 

\section{Conclusions and future work}
\label{sec:conclusions}

In this paper we investigated the correspondence of motion and sound to help visual sound source localization methods. Our proposed method (FlowGrad) and their variations, greatly outperform previous state of the art models for urban sound source localization, showing the importance of motion and temporal context for analyzing urban scenes. For future work, we plan to improve the quality of the optical flow estimation to make it more robust to lighting and camera instability
, and explore the use of multiple frames from as input to the vision encoder as in \cite{senocak2022less}.

\vfill\pagebreak

\bibliographystyle{IEEEbib}
\bibliography{strings,refs}

\end{document}